\documentclass[aps,prl,twocolumn,amssymb,nofootinbib]{revtex4}
\usepackage{epsfig}
\usepackage{amsmath}
\usepackage{amssymb}
\usepackage{graphicx}

\begin{document}
\title{Constraining neutron star tidal Love numbers with gravitational wave detectors}

\author{ \'Eanna \'E. Flanagan}
\author{Tanja Hinderer}
\affiliation{Center for Radiophysics and Space
Research, Cornell University, Ithaca, NY 14853, USA}

\begin{abstract}
Ground-based gravitational wave
detectors may be able to constrain the nuclear equation of state using the
early, low frequency portion of the signal of detected neutron
star - neutron star inspirals. In this early adiabatic
regime, the influence of a neutron star's internal structure on the phase of
the waveform depends only on a single parameter $\lambda$ of the star related to
its tidal Love number, namely the ratio of the induced quadrupole moment to the
perturbing tidal gravitational field. We
analyze the information obtainable from
gravitational wave frequencies smaller than a cutoff frequency of $400{\,}{\rm Hz}$,
where corrections to the internal-structure signal are less than $10
\%$. For an inspiral of two non-spinning $1.4M_{\odot}$ neutron
stars at a distance of $50$ Megaparsecs, LIGO II detectors will be
able to constrain $\lambda$ to $\lambda \leqslant 2.0 \times 10^{37}
{\rm g}{\,}{\rm cm}^2{\rm s}^2 $ with $90\%$ confidence. Fully
relativistic stellar models show that the corresponding constraint
on radius $R$ for $1.4M_{\odot}$ neutron stars would be $R \leqslant
13.6 {\,}{\rm km} {\;}\left(15.3{\,} {\rm km} \right)$ for a $n=0.5$
$\left(n=1.0\right)$ polytrope with equation of state $p \propto
\rho^{1 + 1/n}$.

\end{abstract}

\maketitle

\def\beq{\begin{equation}}
\def\endeq{\end{equation}}
\def\bea{\begin{eqnarray}}
\def\eea{\end{eqnarray}}
\def\bes{\begin{subequations}}
\def\ees{\end{subequations}}

\def\hpl{{{\hat \Phi}_L}}
\def\hpr{{{\hat \Phi}_R}}


\noindent {\it Background and motivation:} Coalescing binary neutron
stars are one of the most important sources for gravitational wave
(GW) detectors \cite{Cutler:2002me}. LIGO I observations have
established upper limits on the event rate \cite{Abbott:2007xi}, and
at design sensitivity LIGO II is expected to detect inspirals at a
rate of $\sim 2/$day
\cite{2004ApJ...601...L179}.

One of the key scientific goals of detecting neutron star (NS)
binaries is to obtain information about the nuclear equation of
state (EoS), which is at present fairly unconstrained in the
relevant density range $\rho \sim 2 - 8 \times 10^{14}{\rm g}{\,}{\rm
cm}^{-3}$ \cite{2000ARNPS..50..481H}. The conventional view has
been that for most of the inspiral, finite-size effects have a
negligible influence on the GW signal, and
that only during the last several orbits and merger at GW frequencies
$f \agt 500$ Hz
can the effect of the internal structure be
seen.

There have been many investigations of how well the EoS can be
constrained using these last several orbits and merger, including
constraints from the GW energy spectrum
\cite{2002PhRvL..89w1102F}, and, for black hole/NS inspirals, from
the NS tidal disruption signal \cite{2000PhRvL..84.3519V}. Several
numerical simulations have studied the dependence of the
GW spectrum on the radius \cite{2003PhR...376...41B}.
However, there are a number of difficulties associated with trying to
extract equation of state information from this late time regime:
(i) The highly complex behavior requires solving the full
nonlinear equations of general relativity together with
relativistic hydrodynamics. (ii) The signal depends on unknown
quantities such as the spins of the stars.
(iii) The signals from the hydrodynamic merger
(at frequencies $\gtrsim $ 1000 Hz) are outside of LIGO's most
sensitive band.

The purpose of this paper is to demonstrate the potential feasibility
of instead
obtaining EoS information from the early, low frequency part of
the signal.  Here, the influence of tidal effects is a small
correction to the waveform's phase, but it is very clean and
depends only on one parameter of the NS -- its Love number \cite{Mora:2003wt}.

\medskip
\noindent {\it Tidal interactions in compact binaries:} The
influence of tidal interactions on the waveform's phase has been
studied previously using various approaches
\cite{1995MNRAS.275..301K,1993ApJ...406L..63L,1992ApJ...398..234K,
1998PhRvD..58h4012T,2002PhRvD..65j4021P,Mora:2003wt}.
We extend those studies by (i) computing the effect of the
tidal interactions for fully relativistic neutron stars, i.e. to
all orders in the strength of internal gravity in each star, (ii)
computing the phase shift analytically without the assumption that the mode
frequency is much larger that the orbital frequency, and (iii)
performing a computation of how accurately the Love number can be
measured.

The basic physical effect is the following:
the $l=2$ fundamental {\it{f}}-modes of the star can
be treated as forced, damped harmonic oscillators driven by the
tidal field of the companion at frequencies below their
resonant frequencies. Assuming circular orbits
they obey equations of motion of the form \cite{Dong94}
\beq
{\ddot q} + \gamma {\dot q} + \omega_0^2 q = A(t) \cos[ m\Phi(t)],
\endeq
where $q(t)$ is the mode amplitude, $\omega_0$ the mode frequency,
$\gamma$ a damping constant, $m$ is the mode azimuthal quantum
number,
 $\Phi(t)$ is the orbital phase of the binary, and $A(t)$ is a
slowly varying amplitude.
The orbital frequency $\omega(t) = {\dot \Phi}$ and $A(t)$ evolve on
the radiation reaction timescale which is much longer than $1/\omega_0$.
In this limit the oscillator evolves
adiabatically, always tracking the minimum of its time-dependent
potential. The energy absorbed by the oscillator up
to time $t$ is
\beq
E(t) = {\omega_0^2A(t)^2 \over 2 (\omega_0^2- m^2 \omega^2)^2 } + \gamma \int_{-\infty}^t dt'
{m^2\omega(t')^2 A(t')^2 \over w_0^4 + m^2\omega(t')^2 \gamma^2}.
\endeq
The second term here describes a cumulative, dissipative effect
which dominates over the first term for tidal interactions of main
sequence stars. For NS-NS binaries, however, this term is
unimportant due to the small viscosity \cite{1992ApJ...398..234K},
and the first, instantaneous term dominates.

The instantaneous effect is somewhat larger than often estimated
for several reasons: (i) The GWs from the time varying stellar
quadrupole are phase coherent with the orbital GWs,
and thus there is a contribution to the energy flux that is linear
in the mode amplitude. This affects the rate of inspiral and gives
a correction of the same order as the energy absorbed by the mode
\cite{1993ApJ...406L..63L}. (ii) Some papers
\cite{1992ApJ...398..234K,1995MNRAS.275..301K,1998PhRvD..58h4012T}
compute the orbital phase error as a function of
orbital radius $r$. This is insufficient as one has to express it
in the end as a function of the observable frequency, and there is
a correction to the radius-frequency relation which comes in at
the same order. (iii) The effect scales as the fifth power of
neutron star radius $R$, and most previous estimates took
$R=10\,{\rm km}$. Larger NS models with e.g. $R = 16{\,} {\rm km}$
give an effect that is larger by a factor of $\sim 10$.

\medskip
\noindent {\it Tidal Love number:} Consider a static, spherically
symmetric star of mass $m$ placed in a time-independent external
quadrupolar tidal field  ${\cal E}_{ij}.$ The star will develop in
response a quadrupole moment $Q_{ij}.$ In the star's local
asymptotic rest frame (asymptotically mass centered Cartesian coordinates) at
large $r$ the metric coefficient $g_{tt}$ is given by (in units
with $G=c=1$) \cite{Thorne:1998kt}:
\beq
\frac{\left(1-g_{tt} \right)}{2} = - {m \over r}- {3Q_{ij} \over 2
r^3} \left( n^i n^j - \frac{\delta^{ij}}{3} \right) +{{\cal
E}_{ij} \over 2} x^i x^j + \ldots
\label{eq:Qdef}
\endeq
where $n^i=x^i/r;$ this expansion defines the traceless tensors ${\cal E}_{ij}$ and
$Q_{ij}.$ To linear order, the induced quadrupole will be of the
form
\beq
Q_{ij} =- \lambda {\cal E}_{ij}.
\label{eq:Lovedef}
\endeq
Here $\lambda$ is a constant which we will call the tidal Love number
(although that name is usually reserved for the dimensionless quantity
$k_2 = {3\over 2}G \lambda R^{-5}$).
The relation (\ref{eq:Lovedef})
between $Q_{ij}$ and ${\cal E}_{ij}$
defines the Love number $\lambda$ for both Newtonian and
relativistic stars.
For a Newtonian star, $ \left(1-g_{tt}
\right)/2$
 is the Newtonian potential, and $Q_{ij}$ is related
to the density perturbation $\delta \rho$  by $Q_{ij}=\int d^3x
\delta \rho \left(x_i x_j -  r^2 \delta_{ij}/3\right)$.

We have calculated the Love numbers for a variety of fully
relativistic NS models with a polytropic pressure-density relation
$P=K \rho^{1+1/n}$.
Most realistic EoS's resemble a polytrope with
effective index in the range $n \simeq 0.5-1.0$
\cite{2001ApJ...550..426L}. The equilibrium
stellar model is obtained by numerical integration of the Tolman-Oppenheimer-Volkhov
equations. We calculate the linear
$l=2$ static perturbations to the Schwarzschild spacetime
following the method of \cite{1967ApJ...149..591T}. The perturbed
Einstein equations $\delta G_{\mu}{}^{\nu}= 8 \pi \delta
T_{\mu}{}^{\nu}$ can be combined into a second order differential
equation for the perturbation to $g_{tt}$. Matching the exterior
solution and its derivative to the asymptotic expansion (\ref{eq:Qdef}) gives
the Love number.
For $m/R \sim 10^{-5}$ our results
agree well with the Newtonian results of Refs.\
\cite{1995MNRAS.275..301K,1955MNRAS.115..101B}.
Figure 1 shows
the range of Love numbers for $m/R=0.2256$, corresponding to the
surface redshift $z=0.35$ that has been measured for
EXO0748-676 {\cite{2002Natur.420...51C}}.
Details of this computation
will be published elsewhere.

\begin{figure}
\begin{center}
\includegraphics*[viewport=1 1 200 270]{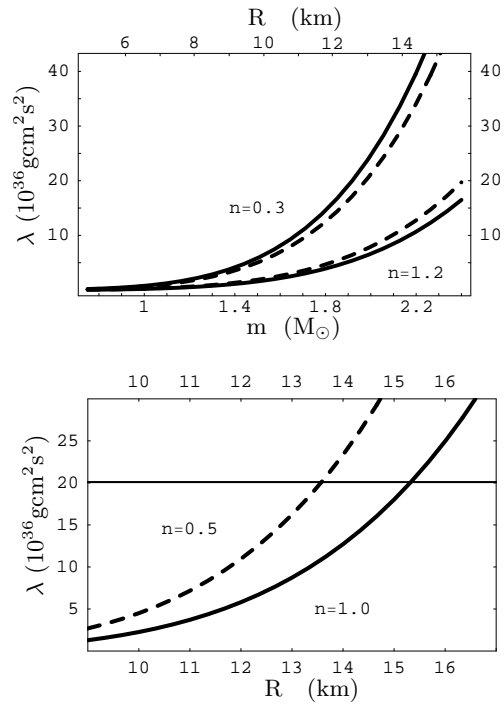}
\caption{[Top] The solid lines bracket the range of Love numbers
$\lambda$ for fully relativistic polytropic neutron star models of
mass $m$ with surface redshift $z=0.35$, assuming a range of $0.3
\leq n \leq 1.2$ for the adiabatic index $n$. The top scale gives
the radius $R$ for these relativistic models. The dashed lines are
corresponding Newtonian values for stars of radius $R$. [Bottom]
Upper bound (horizontal line) on the weighted average ${\tilde
\lambda}$ of the two Love numbers obtainable with LIGO II for a
binary inspiral signal at distance of 50 Mpc, for two non-spinning,
$1.4 M_\odot$ neutron stars, using only data in the frequency band
$f<400$ Hz. The curved lines are the actual values of $\lambda$
for relativistic polytropes with $n=0.5$ (dashed line) and $n=1.0$
(solid line).} \label{fig1}
\end{center}
\end{figure}

\medskip
\noindent {\it Effect on gravitational wave signal:}
Consider a binary with masses $m_1$, $m_2$ and Love numbers
$\lambda_1$, $\lambda_2$.  For simplicity, we compute only the excitation of
star 1; the signals from the two stars simply add in the phase.
Let $\omega_n$, $\lambda_{1,n}$ and $Q_{ij}^n$
be the frequency, the
contribution to $\lambda_1$ and the contribution to $Q_{ij}$
of modes of the star with $l=2$ and with
$n$ radial nodes, so that $\lambda_1 = \Sigma_n \lambda_{1,n}$ and $Q_{ij}
= \Sigma_n Q_{ij}^n$.
Writing the relative displacement as
${\bf x} = (r \cos \Phi, r \sin \Phi, 0)$, the action for the system is
\begin{eqnarray}
S &=& \int dt \left[{1 \over 2} \mu {\dot r}^2 + {1 \over 2} \mu
r^2 {\dot
    \Phi}^2 + { M \mu \over r} \right] -  {1 \over 2} \int dt Q_{ij} {\cal
    E}_{ij} \nonumber \\
&& + \sum_n \int dt {1 \over 4 \lambda_{1,n} \omega_n^2} \left[ {\dot Q}^n_{ij} {\dot
    Q}^n_{ij} - \omega_n^2 Q^n_{ij} Q^n_{ij} \right].
\label{eq:action}
\end{eqnarray}
Here $M$ and $\mu$ are the total and reduced masses, and ${\cal
  E}_{ij} =-m_2 \partial_i \partial_j \left( 1/r \right)
$ is the tidal field.
This action is valid to leading
order in the orbital potential but to all orders in the internal
potentials of the NSs, except that it neglects GW dissipation,
because $Q_{ij}$ and ${\cal E}_{ij}$ are defined in the star's local
asymptotic rest frame \cite{TH}.

Using the action (\ref{eq:action}), adding the leading order,
Burke-Thorne GW dissipation terms, and defining the total quadrupole
$Q_{ij}^{\rm T}= Q_{ij}+ \mu x_i x_j - \mu r^2 \delta_{ij}/3$
with $Q_{ij} = \Sigma_n Q_{ij}^n$, gives the equations of motion
\begin{subequations}
\label{eq:system}
\begin{eqnarray}
{\ddot x^i}  + {M \over r^2} n^i &=& {m_2 \over 2 \mu} Q_{jk}
\partial_i \partial_j \partial_k {1 \over r} - {2 \over 5} x_j {d^5
  Q^{\rm T}_{ij} \over dt^5}, \\
{\ddot Q}^n_{ij} + \omega_n^2 Q^n_{ij} &=& m_2 \lambda_{1,n} \omega_n^2
\partial_i
  \partial_j {1 \over r} - {2 \over 5} \lambda_{1,n} \omega_n^2 {d^5
  Q^{\rm T}_{ij} \over dt^5}.\ \ \ \
\end{eqnarray}
\end{subequations}
By repeatedly
differentiating  $Q_{ij}^{\rm T}$ and eliminating
second order time derivative terms using
the conservative parts of Eqs.\ (\ref{eq:system}), we can
express $d^5 Q^{\rm T}_{ij}/ dt^5$ in terms of $x^i$, ${\dot
x}^i$, $Q^n_{ij}$ and ${\dot Q}^n_{ij}$
and obtain a second order set of equations; this casts Eqs. (\ref{eq:system}) into
a numerically integrable form.

When GW damping is neglected, there exist
equilibrium solutions with $r={\rm const}$,
$\Phi=\Phi_0 + \omega t$ for which $Q_{ij}^{\rm T}$ is static in the
rotating frame. Working to leading order in $\lambda_{1,n}$,
we have
$Q^{\rm T}_{11}= {\cal Q}^\prime + {\cal Q}\cos(2 \Phi)$,
$Q^{\rm T}_{22}= {\cal Q}^\prime - {\cal Q}\cos(2 \Phi)$,
$Q^{\rm T}_{12}= {\cal Q}\sin(2 \Phi)$,
$Q^{\rm T}_{33}= -2 {\cal Q}^\prime$,
where
\beq
{\cal Q} =
\frac{1}{2} \mu r^2 + \sum_n {3 m_2 \lambda_{1,n}
\over 2  (1 - 4 x_n^2)r^3},\ {\cal Q}^\prime=\frac{1}{6} \mu r^2
+ \sum_n { m_2 \lambda_{1,n}
\over 2r^3}
\endeq
and $x_n = \omega / \omega_n$.
Substituting these solutions back into the action (\ref{eq:action}),
and into the quadrupole formula $\dot E=-{1 \over 5} \langle
\dddot Q^{\rm T}_{ij} \dddot Q^{\rm T}_{ij} \rangle$ for the GW damping, provides
an effective description of the orbital dynamics for quasicircular
inspirals in the adiabatic limit.  We obtain for the orbital radius,
energy and energy time derivative
\bes
\bea
r(\omega) &=& M^{1/3} \omega^{-2/3}
\left[ 1 + \frac{3}{4} \sum_n \chi_n
  g_1(x_n) \right], \\
\label{eq:adE}
E(\omega) &=& - { \mu \over 2} \left(M \omega\right)^{2/3}
\left[ 1 - \frac{9}{4} \sum_n \chi_n
  g_2(x_n) \right], \\
{\dot E}(\omega) &=& - \frac{32}{5} M^{4/3} \mu^2 \omega^{10/3}
\left[ 1 + 6 \sum_n \chi_n
  g_3(x_n) \right],\ \ \
\eea
\ees
where $\chi_n = m_2 \lambda_{1,n} \omega^{10/3} m_1^{-1} M^{-5/3}$,
$g_1(x) = 1 + 3 / (1-4x^2)$, $g_2(x) = 1 + (3 - 4x^2) (1 - 4 x^2)^{-2}$,
and $g_3(x) = (M/m_2 + 2 - 2 x^2)/(1-4x^2)$.
Using the formula $d^2 \Psi/d \omega^2 =
2 \left(dE/d\omega\right)/\dot E$ for the phase $\Psi(f)$ of the Fourier
transform of the GW signal at GW frequency $f=\omega/\pi$ {\cite{Tichy}} now gives
for the tidal phase correction
\bea
\label{eq:tidal0}
\delta \Psi(f)&=& - \frac{15m_2^2 }{16 \mu^2
M^5} \sum_n \lambda_{1,n} \int_{v_i}^v dv' v' \left(v^3-v'^3\right)
g_4(x_n'), \nonumber \\
g_4(x) &=&
 \frac{2M}{m_2 (1 - 4 x^2)} + \frac{22 - 117 x^2 +
348 x^4 - 352 x^6}{(1 - 4 x^2)^3}. \nonumber\\
\eea
Here $v = (\pi M f)^{1/3}$, $v_i$ is an arbitrary constant related to
the initial time and phase of the waveform, and $x_n' =
(v')^3/(M\omega_n)$.  In the limit $\omega \ll \omega_n$ assumed in
most previous analyses \cite{1995MNRAS.275..301K,1992ApJ...398..234K,
1998PhRvD..58h4012T,Mora:2003wt},
we get
\begin{equation}
\label{eq:tidal1}
\delta \Psi = - \frac{9}{16} \frac{v^5}{\mu M^{4}}\left[ \left( 11
\frac{m_2}{m_1}+ \frac{M}{m_1} \right) \lambda_1 + 1
\leftrightarrow 2 \right ],
\end{equation}
which depends on internal structure only
through $\lambda_1$ and $\lambda_2$.
Here we have added the contribution from star 2.
The phase (\ref{eq:tidal1}) is formally of post-5-Newtonian (P5N) order, but
it is larger than the point-particle P5N terms (which are currently unknown)
by $\sim (R/M)^5 \sim 10^5$.

\begin{figure}
\begin{center}
\includegraphics[width=.45\textwidth]{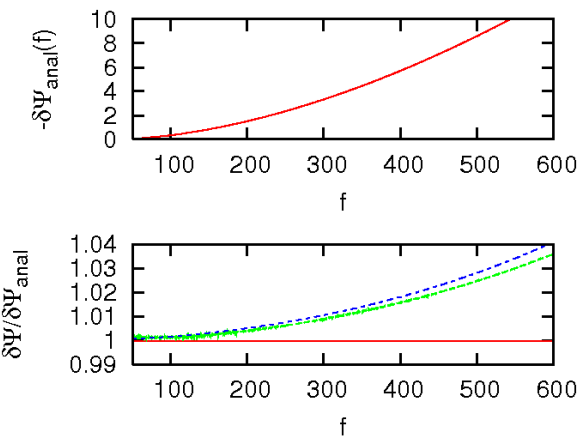}
\caption{[Top] Analytic approximation (\ref{eq:tidal1}) to the
tidal perturbation to the gravitational wave phase
for two identical $1.4 M_\odot$ neutron stars of radius $R = 15\,$km,
modeled as $n=1.0$ polytropes, as a function of
gravitational wave frequency $f$.
[Bottom] A comparison of different approximations to the tidal phase
perturbation: the numerical solution (lower dashed, green curve) to the
system (\ref{eq:system}), and the adiabatic analytic approximation
(\ref{eq:tidal0}) (upper dashed, blue), both
in the limit (\ref{simplify}) and divided by the leading
order approximation (\ref{eq:tidal1}).}
\label{fig2}
\end{center}
\end{figure}

\medskip
\noindent
{\it Accuracy of Model:} We will analyze the information contained in
the portion of the signal before $f = 400 \, {\rm Hz}$.  This
frequency was chosen to be at least $20\%$ smaller than the frequency
of the innermost stable circular orbit \cite{1996PhRvD..54.3958L}
for a conservatively large
polytropic NS model with $n=1.0$, $M = 1.4 M_{\odot}$, and $R = 19\,$km.
We now argue that in this frequency band,
the simple model (\ref{eq:tidal1}) of the phase correction is sufficiently
accurate for our purposes.

We consider six types of corrections to (\ref{eq:tidal1}).
For each correction, we estimate its numerical value at
the frequency $f=400$ Hz for a binary of two identical $m=1.4
M_\odot$, $R = 15$, $n=1.0$ stars: (i)
Corrections due to modes with $l\ge3$ which
are excited by higher order tidal tensors
${\cal E}_{ijk}, \ldots$.  The
$l=3$ correction to $E(\omega)$, computed using the above methods in the low frequency limit,
is smaller than the $l=2$ contribution by a factor of $65 k_3 R^2 /
(45 k_2 r^2)$, where $k_2$, $k_3$ are apsidal constants.
For Newtonian polytropes we have $k_2 = 0.26$, $k_3 = 0.106$
\cite{Mora:2003wt},
and the ratio is
$0.58 (R/r)^2 = 0.04 (R/15\,{\rm km})^{2}$.  (ii)
To assess the accuracy of the $\omega \ll \omega_n$ limit underlying (\ref{eq:tidal1})
we simplify the model (\ref{eq:action}) by taking
\beq
\label{simplify}
\omega_n = \omega_0\ \ \ \ {\rm for}\ {\rm all}\ n,
\endeq
so that $Q^n_{ij}/\lambda_{1,n}$ is independent of $n$.
This simplification does not affect (\ref{eq:tidal1}) and {\it increases} the size
of the finite frequency corrections in (\ref{eq:tidal0}) since
$\omega_n \ge \omega_0$ \footnote{Buoyancy forces and
  associated $g$-modes for which $\omega_n \le \omega_0$ have a
  negligible influence on the waveform's phase\cite{Dong94}.}.  This
will yield an upper bound on the size of the corrections.
(Also the $n\ge1$ modes contribute typically less than $1-2\%$ of the
Love number \cite{1995MNRAS.275..301K}.)
Figure 2 shows the phase correction $\delta \Psi$ computed numerically
from Eqs.\ (\ref{eq:system}), and the approximations (\ref{eq:tidal0})
and (\ref{eq:tidal1}) in the limit (\ref{simplify}).  We see that the adiabatic approximation
(\ref{eq:tidal0}) is extremely accurate, to better than $1\%$, and so the
dominant error is the difference between (\ref{eq:tidal0}) and
(\ref{eq:tidal1}).  The fractional correction to (\ref{eq:tidal1}) is
$\sim 0.7 x^2 \sim 0.2 (f/f_0)^2$, where $f_0 = \omega_0 / (2 \pi)$,
neglecting unobservable terms of the form $\alpha + \beta f$.
This ratio is $\alt 0.03$ for
$f \le 400$ Hz and for $f_0 \ge 1000$ Hz as is the case for $f$-mode
frequencies for most NS models \cite{2002PhRvD..65j4021P}.
(iii) We have linearized in $\lambda_{1}$; the corresponding
fractional corrections scale as $(R/r)^5 \sim 10^{-3} (R/ 15\,{\rm
km})^5$ at 400 Hz.  (iv)
The leading nonlinear hydrodynamic corrections can be computed by
adding a term $- \alpha Q^0_{ij} Q^0_{jk} Q_{ki}^0$ to the Lagrangian
(\ref{eq:action}), where $\alpha$ is a constant.  This corrects the phase shift (\ref{eq:tidal1}) by
a factor $1 - 285 \alpha \lambda_{1,0}^2 \omega^2/968 \sim 0.9995$,
where we have used the models of Ref.\ \cite{Dong1} to estimate $\alpha$.
(v) Fractional corrections to the tidal signal
due to spin scale as $\sim f_{\rm spin}^2/f_{\rm max}^2$,
where $f_{\rm spin}$ is the spin frequency and $f_{\rm max}$ the
maximum allowed spin frequency.  These
can be neglected as
$f_{\rm max} \agt 1000$ Hz for most models and $f_{\rm spin}$ is
expected to be much smaller than this.  (vi) Post-1-Newtonian
corrections to the tidal signal (\ref{eq:tidal1}) will be of order
$\sim M/r \sim 0.05$.  However these corrections will depend only on
$\lambda_1$ when $\omega \ll \omega_n$, and can easily be computed and included in
the data analysis method we suggest here.

Thus, systematic errors in the measured value of
$\lambda$ due to errors in the model should be $\alt 10\%$,
which is small compared to the current uncertainty in
$\lambda$ (see Fig.\ 1).

\medskip
\noindent
{\it Measuring the Love Number:} The binary's parameters
are extracted from the noisy GW signal by integrating the waveform
$h(t)$ against theoretical inspiral templates $h(t,\theta^i)$,
where $\theta^i$ are the parameters of the binary.
The
best-fit parameters $\hat{\theta}^i$ are those that maximize the
overlap integral.
The probability distribution for the signal parameters
for strong signals and Gaussian detector noise is
$p\left(\theta^i
\right)={\cal{N}} {\rm exp}\left(-1/2 {\,}{\Gamma_{ij}\Delta
\theta^i \Delta \theta^j}\right)$ \cite{1994PhRvD..49.2658C},
where $\Delta \theta^i = \theta^i-{\hat \theta}^i$,
$ \Gamma_{ij} = ( \partial h/\partial \theta^i \, , \, \partial
h/\partial \theta^j )$
is the Fisher information matrix,
and the inner product is defined by Eq. (2.4) of Ref.\ \cite{1994PhRvD..49.2658C}.
The rms statistical measurement error in ${\theta}^i$ is then
$\sqrt{\left(\Gamma^{-1}\right)^{ii}}$.

Using the stationary phase approximation and neglecting corrections
to the amplitude, the Fourier transform of the waveform for spinning
point masses is given by $\tilde h(f) = {\cal A} f^{-7/6}
{\rm exp}\left(i\Psi\right)$.  Here the phase $\Psi$ is
\begin{eqnarray}
&&\Psi(f)= 2 \pi f t_c - \phi_c - {\pi \over 4} + {3 M \over 128 \mu}
(\pi M f)^{-5/3}   \nonumber \\
 && \left[1 + {20 \over 9}\left({743 \over 336 } + {11 \over 4}
{\mu \over M } \right)v^{2} -4 (4 \pi-\beta) v^3 \right.   \nonumber \\
&&+  \left. 10 \left( { 3058673 \over 1016064 } + { 5429 \over 1008
} { \mu \over M } + { 617 \over 144 }{ \mu^2\over M^2 } -\sigma
\right)  v^{4} \right.\nonumber\\
&&+\left.\left(\frac{38645\pi}{252}-\frac{65}{3}\frac{\mu}{M}\right)
\ln v\right.\nonumber\\
&&+\left.\left(\frac{11583231236531}{4694215680}-
\frac{640\pi^2}{3}-\frac{6848\gamma}{21}\right)v^6\right.\nonumber\\
&&+\left.
\frac{\mu}{M}\left(\frac{15335597827}{3048192}+\frac{2255\pi^2}{12}+\frac{47324}{63}-
\frac{7948}{9}\right)v^6\right.\nonumber\\
&&+\left.\left(\frac{76055}{1728}\frac{\mu^2}{M^2}-\frac{127825}{1296}\frac{\mu^3}{M^3}
-\frac{6848}{21}\ln (4v)\right)v^6\right.\nonumber\\
&&+\left.\pi\left(\frac{77096675}{254016}+\frac{378515}{1512}\frac{\mu}{M}
-\frac{74045}{756}\frac{\mu^2}{M^2}\right)v^7\right],
\label{eq:phasemodel}
\end{eqnarray}
where $v = (\pi M f)^{1/3}$, $\beta$ and $\sigma$ are spin parameters, and $\gamma$ is
Euler's constant \cite{lrr}. The tidal term (\ref{eq:tidal1}) adds linearly to
this, yielding a phase model with 7 parameters ($t_c,\phi_c,
M,\mu,\beta,\sigma,{\tilde \lambda}$), where $\tilde \lambda= [(11
m_2 + M) \lambda_1/m_1 + (11 m_1 + M) \lambda_2/m_2 ]/26$ is a
weighted average of $\lambda_1$ and $\lambda_2$. We incorporate the
maximum spin constraint for the NSs by assuming a Gaussian prior for
$\beta$ and $\sigma$ as in Ref.~\cite{1994PhRvD..49.2658C}.

Figure 1 [bottom panel] shows the $90\%$ confidence upper limit
${\tilde \lambda} \leqslant 20.1 \times 10^{36}{\,} {\rm g}{\,}{\rm
cm}^2{\rm s}^2 $ we obtain for LIGO II (horizontal line) for two
nonspinning $1.4 M_{\odot}$ NSs at a distance of $50$ Mpc
(signal-to-noise of 95 in the frequency range $20-400$Hz) with
cutoff frequency $f_c=400{\,}{\rm Hz}$, as well as the corresponding
values of $\lambda$ for relativistic polytropes with $n=0.5$ (dashed
curve) and $n=1.0$ (solid line). The corresponding constraint on
radius assuming identical $1.4 M_\odot$ stars would be $R \leqslant
13.6 {\,}{\rm km} {\;}\left(15.3{\,} {\rm km} \right)$ for $n=0.5$
$\left(n=1.0\right)$ polytropes.
Current NS models span the range $10\, {\rm km} \alt R \alt 15 \, {\rm km}$.

Our phasing model (\ref{eq:phasemodel}) is the most accurate available
model, containing terms up to post-3.5-Newtonian (P3.5N) order.
We have experimented with using lower order phase models (P2N, P2.5N,
P3N), and we find that the resulting upper bound on ${\tilde \lambda}$
varies by factors of order $\sim 2$.  Thus there is some associated
systematic uncertainty in our result.  To be conservative, we have
adopted the most pessimistic (largest) upper bound on ${\tilde
  \lambda}$, which is that obtained from the P3.5N waveform.

In conclusion, even if the internal structure signal is too small to
be seen, the analysis method suggested here could start to give
interesting constraints on NS internal structure for nearby events.

This research was supported in part by NSF grants PHY-0140209 and PHY-0457200.
We thank an anonymous referee for helpful comments and suggestions.

\newcommand{\apjl}{Ap. J. Lett.\ }
\newcommand{\apja}{Ap. J.\ }
\newcommand{\aap}{Astron. and Astrophys.}
\newcommand{\cmp}{Commun. Math. Phys.}
\newcommand{\grg}{Gen. Rel. Grav.}
\newcommand{\lr}{Living Reviews in Relativity}
\newcommand{\mnras}{Mon. Not. Roy. Astr. Soc.}
\newcommand{\pr}{Phys. Rev.}
\newcommand{\prsl}{Proc. R. Soc. Lond. A}
\newcommand{\ptrsl}{Phil. Trans. Roy. Soc. London}


\end{document}